## Metaverse'e yatırım yapılır mı? Bilgiyi benimseme modeli bakış açısı ile algılanan riskin satın alma niyetine etkisi

Öğr. Gör. Dr. İbrahim Halil EFENDİOĞLU
Gaziantep Üniversitesi
efendioglu@gantep.edu.tr
ORCID: 0000-0002-4968-375X

**Özet**

Metaverse, fiziksel dünya ile dijital dünyayı birleştiren sanal bir evrendir. İnsanlar bu sanal ortamda oluşturdukları avatarları ile sosyalleşebilmekte, oyun oynayabilmekte hatta alışveriş yapabilmektedir. Yatırım açısından çok hızlı büyüyen Metaverse, tüketiciler için hem kârlı hem de riskli bir alandır. Metaverse'e yatırım amaçlı girebilmek için belirli bir araştırma yapmak ve bilgi toplamak gerekmektedir. Bu doğrultuda çalışmanın amacı, tüketicilerin metaverse dünyası hakkında elde ettiği bilginin kalitesinin, bilginin güvenilirliğinin ve algıladığı riskin, satın alma niyetine etkisini, bilgiyi benimseme modeli bakış açısıyla ile tespit etmektir. Araştırma için metaverse yatırım konusuna ilgi duyan 495 tüketiciden çevrimiçi olarak veri toplanmıştır. Analizlerde AMOS ve SPSS paket programları kullanılmıştır. İlk olarak değişkenlerin temel yapısı için tanımlayıcı istatistiksel analizler yapılmıştır. Daha sonra modelin güvenirliliği ve geçerliliği test edilmiştir. Son olarak önerilen modeli test etmek için yapısal eşitlik modeli kullanılmıştır. Elde edilen bulgulara göre bilginin güvenilirliği ve kalitesi, satın alma niyetini pozitif ve anlamlı olarak etkilemekte iken algılanan risk, satın alma niyetini negatif ve anlamlı olarak etkilemektedir.

**Anahtar Kelimeler:** metaverse, bilgiyi benimseme modeli, algılanan risk, satın alma niyeti

## Can I Invest in Metaverse? The Effect of Obtaıned informatıon And Perceıved Risk on Purchase Intention By The Perspective of The Informatıon Adoption Model

**Abstract**

Metaverse is a virtual universe that combines the physical world and the digital world. People can socialize, play games and even shop with their avatars created in this virtual environment. Metaverse, which is growing very fast in terms of investment, is both a profitable and risky area for consumers. In order to enter the Metaverse for investment purposes, it is necessary to do a certain research and gather information. In this direction, the aim of the study is to determine the effect of the quality of the information obtained by the consumers about the metaverse world, the reliability of the information and the perceived risk, on the purchase intention from the point of view of the information adoption model. For the research, data were collected online from 495 consumers who were interested in metaverse investment. AMOS and SPSS package programs were used in the analysis. First, descriptive statistical analyzes were made for the basic structure of the variables. Then the reliability and validity of the model were tested. Finally, the structural equation model was used to test the proposed model. According to the findings, the reliability and quality of the information affect the purchase intention positively and significantly, while the perceived risk affects the purchase intention negatively and significantly.

**Keywords:** metaverse, information adoption model, perceived risk, purchase intention

## 1. Giriş

Metaverse kelimesi ilk olarak 1992 yılında Neal Stephenson tarafından yazılan Snow Crash adlı bilim kurgu romanında kullanılmıştır. Metaverse, fiziksel dünya ile dijital dünyayı harmanlayan görsel bir dünyadır. Son zamanların şüphesiz en popüler kavramı olan ve kullanıcı merkezli gelişmiş teknolojiler gerektiren Metaverse, sanal bir dünyada kullanıcılara daha gerçekçi bir deneyim ve zengin etkinlikler sunmaktadır (Zhao vd., 2022). Meta, ötesi ve universe ise evren ya da âlem sözcüklerinin birleşiminden oluşmaktadır. Bu sanal alemde kullanıcıların avatarlar ile birbirleriyle iletişim kurabilmekte ve etkileşimde bulunabilmektedir (Çelikkol, 2022). Metaverse, kullanıcıları için avatarlarla temsil edilen farklı etkinliklere izin verilen, sürükleyici ve paylaşılan bir sanal dünya olarak tanımlanır (Vidal-Tomás, 2022). Dünyanın dört bir yanından kullanıcıların gözlük ve kulaklık aracılığıyla erişip bağlanabileceği bilgisayar grafiklerinden oluşturulan paralel bir sanal gerçeklik evrenini temsil etmektedir (Azar vd., 2022).

Metaverse, temelde sosyal medyanın yerini almayacak, bunun yerine yeni kullanıcı deneyimiyle dolu çevrimiçi bir sosyal medya dünyası sunacaktır. Kullanıcılar kendilerine benzeyen ve hareketlerini taklit eden avatarlar aracılığıyla bir araya gelebilirler. Böylece birbirleriyle fiziksel dünyayı kopyalayan bir çevre ile etkileşime girebilirler (Hollensen vd., 2022). Metaverse'ler, her türlü insan-bilgisayar etkileşimini gerçekleştirmek için kullanılması beklenen yeni ara yüzlerdir. Artırılmış gerçeklik, sanal dünyalar gibi teknolojilere dayanan yeni bir paradigmadır. Ancak iş, öğrenim, ticaret veya eğlence alanında metaverse henüz yaygınlaşmamıştır (Prieto vd., 2022). İnsanların bu sanal dünyaya çalışma, sanat, eğlence, yatırım, eğitim, sosyalleşme ve oyun oynama amaçlı katılabilmektedir (Ağırman ve Barakalı, 2022). İnsanlar Metaverse teknolojisine uyum sağladığı takdirde, sanal gerçeklik cihazları sayesinde; alışveriş yapma, sinemaya gitme, kafede zaman geçirme gibi pek çok eylemi fiziksel bir çaba harcamaksızın yapma fırsatına erişecektir. Böylece insanlar sanal ortamda oluşturdukları avatarları ile sosyalleşebilmekte, oyun oynayabilmekte, hatta alışveriş yapabilmektedir.

İşletmeler için metaverse'ün hemen hemen her sektörde yeni bir devrim yaratacağı tahmin edilmektedir. Bu sanal dünyada yeni beceri, meslek ve sertifikalara ihtiyaç duyulacak ve yeni endüstriler, pazarlar ve kaynaklar oluşacaktır. Markalar için ürün veya hizmetlerinin işlevselliğini artacaktır. Örneğin adını "Meta" olarak değiştiren Facebook, gelecekte 3D olarak iş yapılacak bir yer olan meta veri deposunu oluşturmaya çalışmaktadır. Tüm bu değişikliklerin mali değeri trilyonlarca dolar olacaktır (Hollensen vd., 2022). Diğer taraftan Metaverse dijital alandaki tüketicilerin beklentilerini karşılayabilecektir. Bunun nedeni tüketicilerin beğenisine sunulan ürünler metaverse ortamında deneyimlenebilmelidir (Çelikkol, 2022). Böylece tüketicilerle ilişkileri güçlendirmek daha efektif bir biçimde sağlanmış olmaktadır.

Yatırım için Somnium Space, Axie Infinity, Mirandus, Star Atlas, Illuvium, Decentraland, The Sandbox gibi firmalar ön plana çıkmaktadır. 2021'deki metaverse gayrimenkul satışları 500 milyon dolar civarında gerçekleşmiştir. Mevcut eğilimler, satışların 2022'de iki katına çıkabileceği yönünde öngörülüyor (Mileva, 2022). Hatta 2030 yılına kadar 30 trilyon dolara ulaşacağı tahmin ediliyor (Gil, 2021). Ayrıca teknoloji şirketleri Meteverse'ün yaygınlaşması ve kullanımının çeşitlenmesi için çok büyük miktarda yatırımlar yapmaktadır (Ağırman ve Barakalı, 2022).

Henüz literatürdeki araştırmaların çoğu metaverse'in kullanılabilirliğini, fırsatları, tehditleri, avantajlarını ve dezavantajlarını konu almaktadır (Azar vd., 2022; Vidal-Tomás, 2022; Zhao vd., 2022). Ancak tüketici ve yatırım bakış açısı ile yaklaşımlar hem çok yeni hem de çok azınlıktadır (Hollensen vd., 2022; Prieto vd., 2022).

Bilgiyi benimseme modeli, bir kişinin aktarılan bilgiyi benimseme olasılığını ölçerek kullanıcıların bilgi aktarım niyetini açıklamaya çalışır. Daha çok insanların bilgisayar aracılı iletişim platformlarındaki bilgilerden nasıl etkilendiğini açıklamak için kullanılmaktadır (Sussman ve Siegal, 2003). Model geliştirilerek elektronik ağızdan ağıza pazarlama için kullanılmıştır (Erkan ve Evans, 2016). Bu çalışmanın amacı, tüketicilerin metaverse hakkında elde ettiği bilginin güvenilirliğinin, kalitesinin ve risk algısının satın alma niyetine etkisini bilgiyi benimseme modeli bakış açısıyla tespit etmektir.

## 2. Yöntem

Araştırmada nicel araştırma yöntemi kullanılmıştır. Literatür taraması çerçevesinde ve bilgiyi benimseme modeli bakış açısından tüketicilerin metaverse hakkında elde ettiği bilginin kalitesi ile güvenilirliğinin ve risk algısının satın alma niyetine etkisi incelenecektir. Araştırma için metaverse yatırım konusuna ilgi duyan 495 tüketiciden çevrimiçi olarak veri toplanmıştır. Analizlerde AMOS ve SPSS paket programları kullanılmıştır.

Değişkenlerin temel yapısı için tanımlayıcı istatistiksel analizler yapılmıştır. Öncelikle araştırmaya katılan katılımcılara metaverse yatırım yapmayı düşünüp düşünmediği filtre soru olarak sorulmuş ve eğer katılımcı metaverse yatırım yapmak düşüncesi varsa diğer sorulara devam edilmiştir. Formun ilk bölümünde katılımcıların cinsiyet, medeni durum, yaş, meslek, eğitim ve gelir durumuna tespit etmeye yönelik demografik sorular mevcuttur. İkinci bölümde ise edindiği bilginin güvenilirliği, bilginin kalitesi, algıladığı riske ve satın alma niyetine yönelik sorular mevcuttur.

Esas araştırmaya geçilmeden önce anketin uygulanabilirliğini belirlemek amacıyla 44 kişiye pilot uygulama yapılmıştır. Pilot uygulamada anlaşılırlığı, güvenilirliği ve geçerliliği test edilmiştir. Pilot uygulama sonucunda, soruların anlaşıldığı ve anketin analize uygun olduğu değerlendirilmiştir. Araştırmada normallik dağılımını test etmek için çarpıklık ve basıklık değerleri incelenmiştir. Yapılan analiz sonucunda basıklık değerinin -0,451 ile 0,942 değerleri arasında ve çarpıklık değerinin -0,822 ile 1,211 aralığında yer aldığı tespit edilmiştir. Ayrıca Kolmogorov-Smirnov testi normal dağılıma uygun çıkmıştır. Bu durumda çarpıklık ve basıklık değerleri -2 ile +2 arasında olduğundan verilerin normal dağılım gösterdiği kabul edilmiştir (Tabachnick ve Fidell, 2007).

**3. Bulgular**

Demografik özelliklere ait bulgular Tablo 1 'de gösterilmiştir. Katılımcıların 129'su kadın, 366 erkek'sı olmak üzere 495 kişiden oluşmaktadır. Katılımcıların %73,9'u erkektir. Katılımcılardan 294 kişi bekar ve 201 kişi evlidir. Bu durumda katılımcıların çoğunluğu bekârdır. Ayrıca katılımcıların yaş dağılımları incelediğinde çoğunluk 24 yaş altındadır. Bunun yanında katılımcıların büyük bir bölümü öğrencidir. Gelir düzeyleri açısından incelendiğinde çalışanların çoğu 5.000 TL ile 10.000 TL arasında aylık ücrvdmaktadır.

**Tablo 1.** Demografik Verilerin Frekans Analizi

| Demografik Değişken | Kategoriler | N |
|---|---|---|
| **Cinsiyet** | Kadın | 129 |
| | Erkek | 366 |
| **Medeni Durum** | Bekâr | 294 |
| | Evli | 201 |
| **Yaş** | 18-24 | 185 |
| | 25-34 | 148 |
| | 35-44 | 125 |
| | 45-54 | 29 |
| | 55-64 | 7 |
| | 65 ve üstü | 1 |
| **Meslek** | Öğrenci | 195 |
| | Kamu Çalışanı | 94 |
| | Özel Sektör Çalışanı | 99 |
| | Kendi İşi-Esnaf | 68 |
| | Ev Hanımı | 1 |
| | Emekli | 2 |
| | Çalışmıyor | 36 |
| **Eğitim** | İlköğretim | 39 |
| | Lise | 65 |
| | Önlisans | 98 |
| | Lisans | 225 |
|  | Lisansüstü | 68 |
| **Gelir** | 5000 TL ve altı | 112 |
|  | 5001 TL - 10000 TL | 199 |
|  | 10001 TL - 15000 TL | 82 |
|  | 15001 TL - 20000 TL | 74 |
|  | 25001 TL ve üstü | 28 |

Araştırmadan elde edilen verilere ilk olarak doğrulayıcı faktör analizi uygulanmış ve faktör yapısının geçerliliğinin doğrulanıp doğrulanmadığını incelenmiştir. Bunun için faktör yüklerine ve uyum iyiliği değerlerine bakılmıştır. Elde edilen bulgular sonucunda (CMIN/DF=2,766; CFI=0.928; AGFI=0.919; GFI=0.903; RMSEA= 0,04) ölçek maddelerinin ilgili faktörlere kabul edilebilir bir uyum göstererek yüklendiği görülmüştür. Ardından ölçeğin yapı geçerliliğini tespit etmek amacıyla açımlayıcı faktör analizi kullanılmıştır. Faktör analizinin yapılabilmesi için KMO testi (0,819) sonucunun ve Bartlett"s test sonucunun da (0,001) uygun olduğu görülmüştür. Açımlayıcı faktör analizi sonuçlarına göre ifadeler 4 faktöre yüklenmiştir. Toplamda ise açıklayıcılık oranı %62,52 olarak hesaplanmıştır. Bu durumda kullanılan ölçeğin hazırlanma amacına göre dağılım gösterdiği tespit edilmiştir.

Verilerin güvenilirlik analizi için Cronbach's Alpha (CA) katsayısı, yapı güvenilirliği (Composite Reliability-CR) ve açıklanan ortalama varyans (Average Varience Extracted-AVE) incelenmiştir. Ortalama açıklanan varyans değerinin 0,50'nin ve yapı güvenilirliği ve Cronbach's Alpha değerlerinin 0,70'in üzerinde olması beklenmektedir (Fornell ve Larcker, 1981). Analiz sonucunda Cronbach's Alpha değeri 0,71 olmuştur. Ayrıca Tablo 2'de gösterildiği gibi tüm değişkenlerin güvenilirlik değerlerinin şartları sağladığı görülmüştür.

**Tablo 2.** Güvenilirlik Analizi

| **Faktör** | **CA** | **AVE** | **CR** |
|---|---|---|---|
| Bilginin Kalitesi | 0,842 | 0,622 | 0,726 |
| Bilginin Güvenilirliği | 0,821 | 0,610 | 0,744 |
| Algılanan Risk | 0,833 | 0,602 | 0,727 |
| Satın Alma Niyeti | 0,724 | 0,619 | 0,769 |

Araştırmada önerilen modelin uyum indekslerine ilişkin değerlere bakıldığında veriler modele göre iyi uyum sağlamıştır. Bu durum Tablo 3'de gösterilmiştir.

**Tablo 3.** Yapısal Eşitlik Modeli Uyum İyiliği Değerleri

| **Değer** | **İyi Uyum** | **Kabul Edilebilir Uyum** | **Elde Edilen Değerler** |
|---|---|---|---|
| CMIN/DF | ≤ 3 | ≤ 5 | 2,242 |
| CFI | ≥ 0,97 | ≥ 0,90 | 0,915 |
| AGFI | ≥ 0,90 | ≥ 0,85 | 0,886 |
| GFI | ≥ 0,90 | ≥ 0,85 | 0,872 |
| RMSEA | ≤ 0,05 | ≤ 0,08 | 0,04 |

**Kaynak:** Gürbüz ve Şahin, 2016

Yapılan analiz sonucunda elde edilen p anlamlılık değerleri Tablo 4'de gösterilmiştir. Bu sonuçlara göre bilginin kalitesinin (β =.291; p<.05) ve bilginin güvenilirliğinin (β =.334; p<.05) satın alma niyetine pozitif etkisi olduğu; algılanan riskin (β =-.185; p<.05) satın alma niyetine negatif etkisi olduğu görülmüştür. Böylece H1, H2 ve H3 hipotezleri kabul edilmiştir.

Tablo 4.Hipotez Testleri

| Hipotez | Test Edilen Yol | Standardize Reg. | p değeri |
|---|---|---|---|
| H1 | Bilginin Kalitesi -> Satın Alma Niyeti | 0,291 | *** |
| H2 | Bilginin Güvenilirliği -> Satın Alma Niyeti | 0,334 | *** |
| H3 | Algılanan Risk -> Satın Alma Niyeti | -0,185 | **0,03** |

***p<0,01

## 4. Sonuç

Metaverse, sanal gerçeklik ile artırılmış gerçekliği birleştiren, kullanıcıların avatarlarla temsil edildiği ve sanal alanlarda gezindiği sürükleyici bir dünyadır. Gelecekte pazarlama, moda, teknoloji, oyun gibi birçok sektörde önemli etkileri görülmesi beklenmektedir. Fiziksel gerçekliği dijital sanallıkla birleştiren metaverse dünyasına girmeden önce elde edilen bilgi çok önemlidir. Çünkü her teknolojik yenilik ile birlikte fırsatlar ve tehditler birlikte şekillenmektedir. Çalışmada tüketicilerin metaverse yatırımı için elde ettiği bilginin güvenilirliğinin, kalitesinin ve risk algısının satın alma niyetine etkisi incelenmiştir. Bu doğrultuda bilginin kalitesinin ve bilginin güvenilirliğinin satın alma niyetine pozitif etkisi olduğu ancak algılanan riskin negatif etkisi olduğu tespit edilmiştir. Metaverse yatırım ve satın alım açısından hem çeşitli riskler içermekte hem de gelecek vaat etmektedir.

Çalışmada yatırım öncesi güvenilir ve kaliteli bilgi elde etmenin çok önemli olduğu ortaya çıkmıştır. Ayrıca yatırımcıların risk faktörünü detaylıca hesaba kattıklarında satın almaktan vazgeçtiği görülmüştür. Tüketiciler metaverse ortamında satın alma öncesinde uygun durum tespiti yapması ve mümkünse danışmanlık firmalarından yardım alması atılacak doğru bir adım olacaktır. Özellikle güvenilir ve tecrübeli Metaverse yatırımcılarından bilgi alınması faydalı olacaktır. Doğru kaynaklardan elde edinilecek bilgi yatırımcıların satın alma niyetini olumlu etkileyecektir. Pazarlamacılar da nitelikli ve güvenilir iletişimcilerden gelen Metaverse yatırım bilgilerini tüketicilere doğru kanallarla ulaştırmalıdır. Böylece nitelikli bilgi ile beslenen tüketiciler Metaverse yatırımlarını genişletebilecektir. Metaverse işletmeleri ise tüketiciler için faydalı ve güvenilir bilgileri sosyal medyada paylaşması ve bu konuda düzenli eğitimler vermesi satın alımları doğrudan olumlu etkileyecektir.